\documentstyle[12pt]{article}
\begin{document}
\parindent=0pt
\parskip=6pt
\rm

\begin{center}

{\Large \bf
Fluctuation induced first order phase
 transition\\ in thin films of type I superconductors}

\normalsize

\vspace{1cm}

{\large \bf R. Folk$^{a}$},
 {\large \bf D. V. Shopova$^{b, \ast}$},
 {\large \bf  D. I. Uzunov$^{a-c}$}

$^{a}${\em Institut f\"{u}r Theoretische Physik,
 Johannes Kepler Universit\"{a}t Linz, A-4040 Linz, Austria}\\
$^{b}${\em CPCM Laboratory, G. Nadjakov Institute of Solid
 State Physics,\\
 Bulgarian Academy of Sciences, BG-1784 Sofia, Bulgaria} \\
 $^{c}${\em Max-Planck-Institut f\"{u}r Komplexer Systeme,
 01187 Dresden, Germany}

 $^{\ast}${\small Corresponding author: sho@issp.bas.bg}

\end{center}

\vspace{0.5cm}


\vspace{0.5cm}


\vspace{0.5cm}

\vspace{1cm}

\begin{abstract}

The effect of fluctuation induced weakly first order phase transition
 known for three dimensional (3$D$) type I superconductors appears in
 a modified and strongly enhanced variant in thin (quasi-$2D$)
superconducting films. The unusual thermodynamic properties of this
new type of first order phase transitions and the possibility for an
 experimental verification of the effect are established and discussed.

\end{abstract}

\vspace{1cm}

One of the outstanding problems
 of the phase transition theory of superconductivity
 is the so-called fluctuation induced weakly first order phase transition (WFOT)
in three dimensional (3D) type I superconductors. It was
 shown~\cite{hlm:prl} that the interaction between
 the fluctuations ${\cal{\bf \delta H}}$ of
 the magnetic field ${\cal{\bf H}}$ and the order parameter
 $\psi(\cal{\bf x})$ may produce a fluctuation induced
 change of the order of the phase transition in 3D type I
 superconductors: from
a second order phase transition when the magnetic fluctuations
 $\delta \cal{\bf H}$ are
 neglected to a weak first order transition when the same magnetic
 fluctuations
 are taken into account and the mean magnetic field
 ${\cal{\bf H}}_0 =
({\cal{\bf H}} - \delta {\cal{\bf H}})$ is equal to zero. Note, that we
 consider nonmagnetic superconductors,
 where the magnetic induction
${\cal{\bf B}}$
 is equal to the magnetic field ${\cal{\bf H}}$. Besides, we
 discuss the phase transition point $T_{c0}$ corresponding to a zero
 critical magnetic field, $H_c(T_{c0}) = 0$, and our consideration
 will be not extended to the entire line
  [$0 \leq H_c(T) \leq H_c(0)$] of the superconducting phase transition.

       The same type of fluctuation induced first order phase
 transition was predicted also in other fields of physics,
 for example, in the theory of the
 early universe~\cite{Linde:1979, Vil:1994}
 and on the basis of the de Gennes model of smectic A -- nematic phase
 transitions in liquid crystals~\cite{Gennes:1993, Lub:1978}. It seems
 that under certain circumstances,
 i.e. suitable values of the characteristic lengths,
 such fluctuation induced first order
 transitions may
 occur in any system described by an Abelian-Higgs type of
 model~\cite{Linde:1979, Vil:1994, Col:Wei},
 where a scalar field like the
superconducting order parameter $\psi(\cal{\bf x})$  interacts
 with a massless
vector gauge field. In a superconductor the massless gauge field is
 the vector potential ${\cal{\bf A}}({\cal{\bf x)}}$ of the magnetic
 field given for 3D systems by ${\cal{\bf H}}({\cal{\bf x}}) =
 \bigtriangledown \times {\cal{\bf A}}$
 and the Coulomb gauge  $\mbox{div}{\cal{\bf A}} = 0$.
 A general description of the phase transition mechanism in Abelian-Higgs
 systems
 was given within the
 the scalar field electrodynamics~\cite{Col:Wei}. The theoretical
 studies of this problem in superconductors have been performed
 with the help of the Ginzburg-Landau (GL) theory of superconductivity
 and the so-called dual model~\cite{Kio:1994, Kio:1995}),
including  the application of the renormalization group (for a review,
 see~\cite{Uzun:1993, Folk:1999}).
The WFOT problem seems
 to be of general interest to a wide range of systems and, in particular,
 to recent aspects of superconductivity~\cite{Her:1998, Tes:1999}.
 According to the theoretical predictions the effect in 3D superconductors
 is very weak~\cite{hlm:prl, Chen:1978}. For this reason
an experimental search of the WFOT was attempted in smectic A
 liquid crystals
 by interface velocity measurements as well as by
high-resolution heat capacity and X-ray
 experiments~\cite{Ani:1990, Gar:1990}. The results were
 interpreted~\cite{Ani:1990, Gar:1990}
 in favour of a WFOT due to the interaction between
 the smectic scalar order parameter and
 the director vector of the nematic order, i.e. as a verification of the
theoretically established fluctuation mechanism.

We shall restrict our consideration
 to type I superconductors where the description of the WFOT
 can be given within the "approximation of spatially
 uniform order parameter"
($\psi(\cal{\bf x}) \approx \psi$ ~\cite{hlm:prl, Chen:1978}).
That is why we shall present a form
 of the GL free energy which can be
 used for the investigation of the WFOT in D-dimensional
 type I superconductors for a uniform order parameter $\psi$.
The 3D and 4D dimensionalities were analyzed in preceding
 works~\cite{hlm:prl, Chen:1978}
 and we shall focus on quasi-2D films.
 We shall  demonstrate that the effect of WFOT
 is much stronger in films than in 3D samples and, therefore,
 could be experimentally observed (if it really exists). Besides, in
quasi-2D superconductors the WFOT is described by a logarithmic
 term instead of a third order term in $\psi$~\cite{hlm:prl, Uzun:1993},
 and this presents
 a special theoretical interest.

In standard notations~\cite{Lifs:1980} the GL free energy
 in case of a spatially uniform order parameter $\psi$
 takes the form $F_{GL} = (F_0 + F_A)$ with
\begin{equation}
\label{eq1}
F_{0} = V \left ( a|\psi|^2 + \frac{b}{2}|\psi|^4 \right ) \;,
\end{equation}
and
\begin{equation}
\label{eq2}
F_{A} = \frac{1}{16\pi}\sum_{ij}\int d^Dx \left [ 2\rho(\psi)\delta_{ij}A^2_j
 + (\partial_i A_j
 - \partial_jA_i)^2 \right] \;,
\end{equation}
where $\partial_i = \partial/\partial_i$, $(i,j) = 1,...,D$;
 $V = (L_1...L_D)$ is the volume of the D-dimensional superconductor,
$H =|{\cal{\bf H}}|$, $a = \alpha_0 (T - T_{c0})$ with $\alpha_0(T_{c0}) > 0$,
 $b(T_{c0}) > 0$, $\rho = \rho_0 |\psi|^2$ with
 $\rho_0 = (8 \pi e^2/mc^2$). The quantities
 $2m$ and $2e$ are the
 effective mass and charge of the Cooper pairs respectively, and
 in superconductors of interest to the present investigation
 $2m \approx 2m_e$, where $m_e$ is the electron rest mass.
 The concrete values of
the initial
 (bare) critical temperature $T_{c0}$
 and the Landau parameters $\alpha_0$ and $b$
  given by the microscopic BCS theory~\cite{Lifs:1980},
 are not essential for
 our consideration
 within the general GL phenomenological approach.
 But we should keep in mind
 that the GL
 theory  describes only quasimacroscopic phenomena
 with characteristic lengths greater than the zero-temperature
 coherence length $\xi_0 = (\hbar^2/4m\alpha_0T_{c0})^{1/2}$; for a
 microscopic (BCS) justification of this argument,
 see, e.g.,~\cite{Lifs:1980}.

In the space of the wave vectors
 ${\cal{\bf k}}$, the upper cutoff $\Lambda$ for $k \equiv {\cal{\bf k}}$
 is $\Lambda \sim (1/\xi_0)$ and can be extended to larger values only
 in case of special circumstances
 (calculation of cutoff independent integrals). Besides,
 the free energy (1) has additional restrictions which
 are also relevant to our consideration. Firstly,
 the approximation $\psi=$const will be valid for well established
 type I superconductors, like Al, where the GL parameter $\kappa$ is much
 less than unity;
 $\kappa = \lambda(T)/\xi(T)$.
 The coherence length $\xi(T)$ is given by
 $\xi(T) = \xi_0/|t|^{1/2}$ , where $t = (T-T_{c0})/T_{c0}$,
 and the London penetration depth is $\lambda(T) = \lambda_0/|t|^{1/2}$,
 where $\lambda_0 = (mc^2b/8\pi e^2 \alpha_0 T_{c0})^{1/2}$
 stands for the zero-temperature value of $\lambda$. Secondly, we have to take
 into account the additional restriction on the GL theory for type
 I superconductors, namely,
 $|T-T_{c0}| \ll \kappa^{2}T_{c0} $~\cite{Lifs:1980}, which comes
 from the general Landau condition $|T-T_{c0}| \ll T_{c0}$ and
 the requirement for a locality of the functional dependence
 between the supercurrent and the vector potential.
 The latter requirement is fulfilled for slow variations
of the magnetic field and the vector potential at distances
 of order of $\xi_0$,
 i.e. for $\lambda(T) \gg \xi_0$. This means that the GL theory
 for well established type I superconductors with, for example,
 a GL parameter $\kappa \sim 10^{-1} \div 10^{-2}$, is valid only
 for $|t| < 10^{-2} \div 10^{-4}$K. The Ginzburg critical region for
 these superconductors is very small
( $\sim 10^{-12}\div 10^{-16}$K~\cite{Uzun:1993, Lifs:1980}) and can
 be safely ignored. All phenomena that can be
 reliably predicted within this theory are confined to a relatively
 narrow vicinity of the phase transition point ($0 \sim 10^{-12} < |t| <
  10^{-2} \div 10^{-4} $).

Thus we shall use
 the free energy (1) only for the description of phenomena in the
 temperature domain defined by the inequalities
 $\xi(T) \gg \lambda(T) \gg \xi_0$ which are fulfilled
 in the same vicinity of the phase transition point. Our heuristic
 arguments presented above have
 a reliable justification within the microscopic approach ~\cite{Lifs:1980}.

Following preceding papers~\cite{hlm:prl, Col:Wei, Lov:1980}
 we shall
 integrate out the vector potential in order to obtain an effective free
energy as a function of the uniform field $\psi$. This
 integration is performed exactly within a simple one-loop
 expansion, as depicted in Fig.1, where the $\rho-$dependent term in eq. (2)
 is considered as a perturbation. The summation of the infinite
 logarithmic series shown in Fig. 1 yields
\begin{equation}
\label{eq3}
F_{\mbox{eff}}^{(A)} = \frac{1}{2}(D-1)k_B T\sum_{{\cal{\bf k}}} \mbox{ln}
 \left [1 + \rho(\psi)/k^2 \right ]  \;,
\end{equation}
which substitutes $F_A$ in $F_{GL}$. The total effective free energy density is given by
$f_{\mbox{eff}} = (F_0 + F_{\mbox{eff}}^{(A)})/V$. The function
 $f_{\mbox{eff}}(\psi)$ was investigated
in details for $D \geq 3 $ ~\cite{hlm:prl, Chen:1978} when the effects
 of the term $F_{\mbox{eff}}^{(A)}$ are small;
 a brief notice~\cite{Lov:1980} about the
 $2D$ case is also known.

Let us focus the attention on the quasi-2D
 spatial
 dimensionality (3D thin films), where the film thickness $L_0$ satisfies
the condition $ \lambda(T) \gg L_0 \gg \xi_0$. This condition
 is suitable
 for both theoretical and experimental investigations.
Then our relatively thin superconducting slab must obey
 the following conditions:
\begin{equation}
\label{eq4}
 \xi(T) \gg \lambda(T) \gg L_0 \gg \xi_0 \:.
\end{equation}
It is easy to see that the ${\cal{\bf k}}-$summation in eq. (3)
 can be substituted with
 a 2D integration over the wave vector components
 parallel to the film surface and  the transverse
 wave vector component is equal to zero.
 An upper cutoff
 $\Lambda = (1/\xi_0)$ is assumed for $k \equiv |{\cal{\bf k}}|$.
 In contrast to the pure $2D$ case~\cite{Lov:1980} and studies in
 other spatial dimensionalities~\cite{hlm:prl, Col:Wei, Chen:1978}
 the result for the effective free energy of the quasi-$2D$
 superconductor is given by
\begin{equation}
\label{eq5}
f_{\mbox{eff}}  = r_0|\psi|^2 + \frac{u}{2}|\psi|^4 -
Q|\psi|^2\mbox{ln}(\rho_0|\psi|^2/\Lambda^2)  \;,
\end{equation}
where $r_0 = (a + k_B T\rho_0/4\pi L_0)$,
 $u = (b + k_B T\rho_0^2/4\pi \Lambda^2 L_0)$, and
$Q = (k_B T\rho_0/4\pi L_0)$.
The evaluation of the contribution of the vector
 potential fluctuations, i.e. of the $\rho_0$-dependent
 terms in eq. (5) shows that these terms cannot be neglected
for the entire temperature interval of validity
 of our consideration. Besides, the evaluation of
 the new terms in the free energy
is useful as a demonstration of
 the important role of the cutoff $\Lambda$ and
 the necessity of
 the consistent choice $\Lambda \approx (1/\xi_0)$.
 Another choice of $\Lambda$ immediately leads
 to wrong predictions.

    The free energy
 $f(\varphi) = [f_{\mbox{eff}}(\varphi)/u]$
with $\varphi \equiv |\psi| \geq 0$ is depicted in Fig. 2
 for several values of $r = [r_0 + Q\mbox{ln}(\Lambda^2/\rho_0) /u]$,
 and $q = (Q/u)$. Figure 2
 shows a well established first order phase transition.
We shall briefly summarize the analytical treatment of the free energy (5)
 in terms of the simple notations $f$, $r$, and $q$.
The analysis is similar to that for first order phase transitions
 described by
a third order ($\varphi^3-$)term; for details, see ~\cite{Uzun:1993}.
 Let us denote the solutions of the equation of state
($\partial f/\partial \varphi) = 0$ by $\varphi_0 (\equiv |\psi_0|)$
 - the possible phases.
 The normal phase solution
 ($\varphi_0 = 0$) exists and gives
 a minimum of $f$ for all $T \geq 0$. Moreover, the
 necessary stability condition
 $f''(\varphi_0) =(\partial^2 f/\partial^2\varphi)_{\varphi_0} > 0$
 exhibits a singular behaviour
$f''(\varphi_0 \rightarrow 0) \longrightarrow (+\infty)$, i.e. a logarithmic
 divergence.
 This peculiar property
  does not create troubles about the physical meaning of
 the effective free energy. Rather this result shows that in thin films
 the normal state can occur at any $T \geq 0$
 as a stable phase above the equilibrium phase transition temperature
 $T_{c} \neq T_{c0}$, or, if certain experimental conditions are satisfied,
  as a metastable phase for $0 \leq T \leq T_{c}$.

   The superconducting
 (Meissner)
 phase $\varphi_0 > 0$ is described by the equation
\begin{equation}
\label{eq6}
 (r-q) + \varphi_0^2 - q\mbox{ln}\varphi_{0}^{2} = 0  \;.
\end{equation}
The solution of this equation, i.e. the equilibrium
 order parameter $\varphi_0 > 0$
 is shown in Fig. 3 for values of the
 parameters $r$ and $q$ which allow  the
 existence of a first order transition. The check of the necessary
 ($f''(\varphi_0) > 0$) and the sufficient  ($f(\varphi_0) < 0$)
stability
 conditions for $\varphi_0 > 0$ show that the former is fulfilled
 for $\varphi_0^2 > q$, whereas the latter condition is fulfilled for
 $\varphi_0^2 > 2q$. Therefore, the Meissner phase can be overheated above
the equilibrium transition temperature $T_{c}$ of the phase transition defined
 by the equation $\varphi_0^2(T_{c}) = 2q(T_{c})$.
 The equilibrium order parameter jump is equal to $(2q_c)^{1/2}$,
 where $q_c \equiv q(T_c)$.
 This overheating may continue
 up to the temperature $T_{c1}$ defined by
$\varphi_0^2(T_{c1}) = q(T_{c1})$ ,
 where the order parameter becomes equal
 to $q_c^{1/2}$ -- the lowest nonzero value of the order parameter,
 for which it gives a minimum of $f$.
 These features of the order parameter are shown in Fig. 3, where the
 "metastability extension" of the order parameter profile is also
 drawn for a suitable choice of the parameter $q$.
 Note, that
  one can easily show in an analytical way that
 stable solutions $0 < \varphi_0 < 1$
 of the eq. (6) always exist, at least, for $r < 0$.

The latent heat $L(T_{c}) = T_{c}\Delta S(T_{c})$
 and the specific heat jump
 $\Delta C(T_{c}) = T_{c}(\partial S/\partial T)_{T_c}$ at
 the equilibrium phase transition point $T_{c}$ can be
 calculated from the free energy (5). We shall be interested in the
 temperature
{\em size} of this first order transition, namely, in the ratio
$(\Delta T)_{c} = |L(T_{c})/C(T_{c})|$. Taking into account only the
temperature dependence of the effective free energy $f_{\mbox{eff}}$ on
 the Landau parameter $a$, the temperature size $(\Delta T)_{c}$ can be
 correctly evaluated. Our result is
\begin{equation}
\label{eq7}
 (\Delta T)_{c} = \frac{Q}{\alpha_0}  \;,
\end{equation}
and after the  substitution $\alpha_0 = (\hbar^2/4mT_{c0}\xi_0^2)^{1/2}$
 and $Q = (k_B T\rho_0/4\pi \Lambda^2 L_0)$ we have
\begin{equation}
\label{eq8}
 (\Delta T)_{c} = 8k_BT_c^2
\left ( \frac{\xi_0^2}{L_0} \right )\left ( \frac{e}{\hbar c} \right )^2  \;.
\end{equation}

When we put in the above expression the numerical values of the  fundamental
constants and the tabulated data
for Al  ($T_c = 1.19$ K, and $\xi(0) = 1.6 \times 10^{-4}$ cm ~\cite{hlm:prl}),
 we obtain that $ (\Delta T)_{c} \sim 10^{-6}/L_0$. Note, that
 our approach is valid, if
the thickness $L_0$ is greater than $\xi_0$
 and less than $\lambda(T)$. A choice of $L_0$ consistent with all
 theoretical requirements is $ L_0 \sim 10 \times \xi_0$, or, for Al,
 $L_0  \sim 10^{-3}$ cm,
 which yields $ (\Delta T)_{c} \sim 10^{-3}$ K. As the usual GL theory
 for Al is restricted within a temperature interval of size $10^{-4}$ K,
 our result indicates that the size of the WFOT covers totally the
domain of validity of our unusual theory, namely we are faced with
a normal size FOT rather than with a WFOT. Therefore, it should be
emphasized that if the gauge mechanism of a fluctuation induction of
 the phase transitions order change really exists,
 this should be seen in suitable experiments
with pure samples of good type I superconducting films of thickness
about $10\mu$m. The same result for the temperature size of the FOT
 can be obtained by the estimation of the parameter $\alpha_0$ given by the
BCS theory~\cite{Lifs:1980}.

In conclusion, we have three notes:

(1) Our approach to quasi-2D superconductors allows  to use the 3D
 values of the original
GL parameters $\alpha_0$ and $b$ and keeps the consideration far from
dangerous effects of distruction of the order by 2D fluctuations.
In this respect as well as in the values of the effective parameters
 $r_0$, $u$, and $Q$, our effective free energy (5) is quite different
 from the pure
 2D free energy known from a preceding paper of Lovesey~\cite{Lov:1980}.
 That is why we are
 able to make the prediction about the enhancement of the WFOT in
 quasi-2D superconductors (or, which is the same, "almost
 3D" superconductors). We should stress that for pure 2D case
 the Landau parameters $\alpha_0$ and $b$ are different from those
 corresponding to $3D$ superconductors and, hence,
 the evaluation of the same FOT effect in
two dimensions  needs a consistent treatment of these parameters.
 Moreover,
 the investigation of very thin (almost 2D) films requires an evaluation of
 the temperature region where the order parameter fluctuations
 will destroy the phase transition.

For these reasons our results cannot be extended in a straightforward
way to very thin films by the respective decrease of the film thickness $L_0$.
Within the phenomenological approach to dimensional crossover phenomena
~\cite{Suz:Mas}, the 3D Landau parameters are related to the 2D ones by
 $X_{(2D)} = \mbox{lim}_{L_0 \rightarrow 0} (XL_0)$,
 where $X = (\alpha_0, b)$. This consideration should be compared with
the BCS predictions for the 2D values of the free energy parameters.

(2) We have followed a method
 of integration of the magnetic fluctuations proposed
 in preceding papers~\cite{hlm:prl, Col:Wei},
 in which the order parameter fluctuations are not
 totally ignored and the "photon mass" $ \rho(\psi)$ generated by the
effect of the Higgs field $\psi$ on the vector potential ${\cal{\bf A}}$
is not an equilibrium quantity. The special
 feature of the present method is that
the uniform fluctuation $\delta \psi$ of the uniform "field" $\psi$
 interacts with the
magnetic fluctuations and this interaction does ensure
 the fluctuation mechanism of the
FOT transition discussed in the present and all preceding works.
 The WFOT in higher spatial dimensionalities and the FOT in our case
 are caused by
 a high-order fluctuation interaction [$(\delta\psi)^2(\delta A_j)^2$]
 between the order parameter
 and the magnetic field fluctuations rather than
 the magnetic fluctuations alone.

(3) The theoretical arguments in favour of the FOT
 within the approximation $\psi = \mbox{const}$
 below the equilibrium transition point are not precisely
 the same as those obtained by various
 renormalization group investigations in one loop~\cite{hlm:prl, Chen:1978,
Ber:1996} and
higher-order~\cite{Kol:1990, Folk:1996, Folk:1999} approximations
 in the loop expansion, where
 another type of singularities of the perturbation series
 for the GL free energy are relevant. Within the present
 approximation for $\psi$ the divergent terms, which are usually
 relevant for the renormalization group studies and are present in all
 diagrammes in Fig. 1, have been summed up to a finite sum. Therefore,
 the arguments
in favour (or against) a WFOT within the approximation of uniform
 scalar field $\psi$ cannot be used to justify or reject
 renormalization group predictions above the equilibrium transition point.

{\bf Acknowledgments:}

The authors thank the OSI (Vienna and its extension in Sofia) for
 a research grant. DIU thanks the hospitality
 of the Johannes Kepler Universit\"{a}t (Linz) and
 the Max-Planck-Institut f\"{u}r Komplexer Systeme
 (Dresden).

\newpage

{\bf Figure captions}

\vspace{0.5cm}

Fig. 1. One loop logarithmic series of diagrammes: the black squares denote
the vertex part of the $A_j^2-$ term, and the solid lines denote the
correlation function $<|A_j({\cal{\bf k}})|^2>$.

\vspace{0.5cm}

Fig. 2. The shape of the effective free energy $f$
 for $q = 10^{-4}$: line (1)
 corresponds to $r = - 9.3 \times 10^{-4}$, (2) -- to
  $r = - 9.4 \times 10^{-4}$, (3) --
 to  $r = - 9.515 \times 10^{-4}$, and (4) -- to  $r = - 9.7 \times 10^{-4}$.

\vspace{0.5cm}

Fig. 3. The order parameter profile for $q = 10^{-4}$. The black squares indicate the stable
 superconducting states and the white squares indicate metastable superconducting states
above the equilibrium transition point.


\newpage

\begin{thebibliography}{99}


\bibitem{hlm:prl}
B.I.Halperin, T.C.Lubensky, and S.K.Ma,
 Phys. Rev. Lett. 32 (1974) 292.

\bibitem{Linde:1979}
A.D.Linde, Rep. Progr. Phys. 42 (1979) 389.

\bibitem{Vil:1994}
A.Vilenkin, and E.P.S.Shellard,
 Cosmic Strings and Other Topological Defects
 (Cambridge University Press, Cambridge, 1994), ch. 2.


\bibitem{Gennes:1993}
P.G.de Gennes and J.Prost,
 The Physics of Liquid Crystals, 2nd ed. (Clarendon Press, Oxford,1993)



\bibitem{Col:Wei}
S.Coleman and E.Weinberg, Phys. Rev. D7 (1973) 1888.


\bibitem{Lub:1978}
T.C.Lubensky and J-H. Chen, Phys. Rev. B17 (1978) 366.

\bibitem{Chen:1978}
 J-H.Chen, T.C.Lubensky, and D.R.Nelson, Phys. Rev. B17 (1978) 4274.

\bibitem{Uzun:1993}
 D. I. Uzunov, Theory of Critical
 Phenomena
(World Scientific, Singapore, 1993).

\bibitem{Folk:1999}
 R.Folk and Yu.Holovatch, in: Correlations, Coherence, and Order,
 ed. by D. V. Shopova and D. I. Uzunov
 (Kluwer Academic/Plenum Publishers, New York-London, 1999), p.83.


\bibitem{Kio:1994}
 M. Kiometzis, H. Kleinert, and A.M.J.Schakel,
 Phys. Rev. Lett. 73 (1994) 1975.

\bibitem{Kio:1995}
 M. Kiometzis, H. Kleinert, and A.M.J.Schakel,
Forschr. Phys. 43 (1995) 697.

\bibitem{Her:1998}
 I. Herbut, Phys. Rev. B57 (1998) 13729.

\bibitem{Tes:1999}
 Z.Tesanovic, Phys. Rev. B59 (1978) 6449.

\bibitem{Ani:1990}
M.A.Anisimov, P.E.Cladis, E.E.Gorodetskii,
D.A.Huse, V.E.Podneks, V.G.Taratuta, W.van Saarloos,
 and V.P.Voronov, Phys. Rev. A41 (1990) 6449.

\bibitem{Gar:1990}
 C.W.Garland and G.Nounesis, Phys. Rev. E49 (1994) 2964.

\bibitem{Lifs:1980}
 E.M.Lifshitz and L.P.Pitaevskii, Statistical Physics, Part 2,
[Landau and Lifshitz Course of Theoretical Physics, vol. 9]
 (Pergamon Press, Oxford, 1980).

\bibitem{Lov:1980}
 S.W.Lovesey, Z.Physik B - Cond. Matter 40 (1980) 117.

\bibitem{Suz:Mas}
M.Suzuki and D.I.Uzunov, Physica A216 (1995) 1347.

\bibitem{Ber:1996}
B.Bergerhoff, F.Freire, D.F.Litim, S.Lola,
and C.Wetterich, Phys. Rev. B53 (1996) 5734.

\bibitem{Kol:1990}
S. Kolnberger and R. Folk, Phys. Rev. B41 (1990) 4083.
 .

\bibitem{Folk:1996}
R. Folk and Yu. Holovatch, J. Phys. A: Math. Gen. 29 (1996) 3409.

\end{thebibliography}
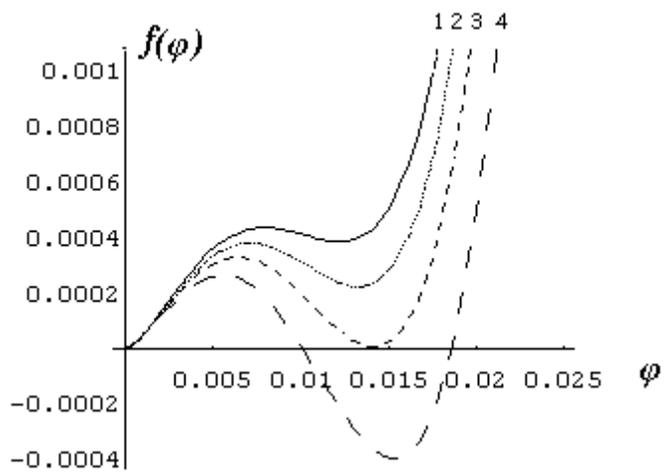
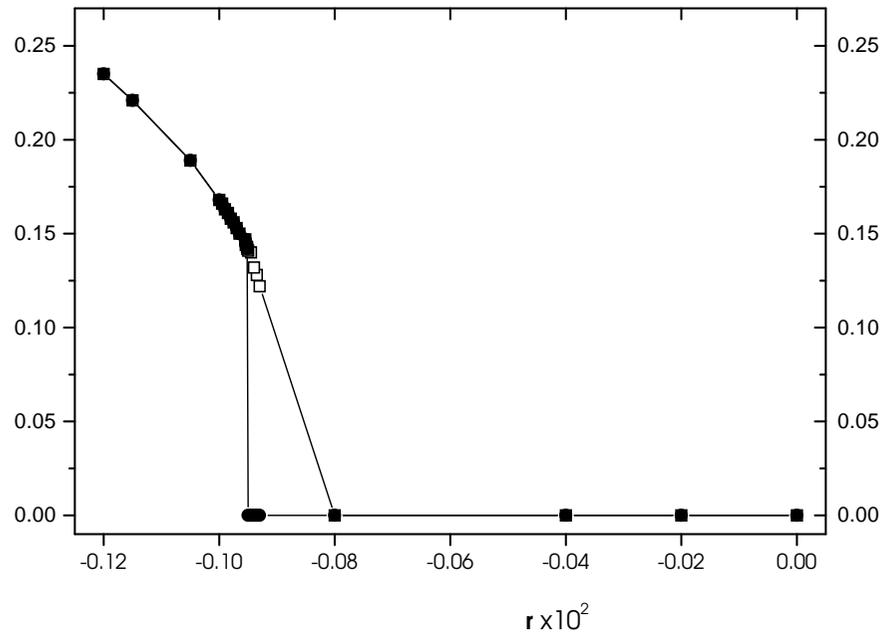
\end{document}